%% file: main-ieee.tex
\documentclass[conference]{IEEEtran}
\IEEEoverridecommandlockouts
\usepackage{cite}
\usepackage{amsmath,amssymb,amsfonts}
\usepackage{algorithmic}
\usepackage{graphicx}
\usepackage{textcomp}
\usepackage{xcolor}
\usepackage{hyperref}
\def\BibTeX{{\rm B\kern-.05em{\sc i\kern-.025em b}\kern-.08em
    T\kern-.1667em\lower.7ex\hbox{E}\kern-.125emX}}
\usepackage{enumitem}
\usepackage{booktabs}
\usepackage{adjustbox}
\usepackage{colortbl}
\usepackage{float}
\usepackage[most]{tcolorbox}
\usepackage{multirow,tabularx}
\newcommand{\nd}{\vspace{1mm}\noindent}
\usepackage{amsmath}
\usepackage{balance}
\usepackage{paralist}

\usepackage{amsfonts}
\usepackage{amssymb}
\makeatletter

\newcommand*\circled[1]{\tikz[baseline=(char.base)]{
            \node[shape=circle,draw,inner sep=2pt] (char) {#1};}}
            
\newcommand*{\rom}[1]{\expandafter\@slowromancap\romannumeral #1@}
\makeatother

\begin{document}

\title{An Empirical Study on the Usage of Automated Machine Learning Tools\\
%
}

\author{\IEEEauthorblockN{Forough Majidi\textsuperscript{\dag}, Moses Openja\textsuperscript{\dag}, Foutse Khomh\textsuperscript{\dag}, Heng Li\textsuperscript{\dag}}
\IEEEauthorblockA{
\{forough.majidi, openja.moses, foutse.khomh, heng.li\}@polymtl.ca}

\IEEEauthorblockA{\textsuperscript{\dag}Polytechnique Montréal, Montréal, Quebec, Canada 
}
}


\maketitle

\begin{abstract}
The popularity of automated machine learning (AutoML) tools in different domains has increased over the past few years. Machine learning (ML)  practitioners use AutoML tools to automate and optimize the process of feature engineering, model training, and hyperparameter optimization and so on. Recent work performed qualitative studies on practitioners' experiences of using AutoML tools and compared different AutoML tools based on their performance and provided features, but none of the existing work studied the practices of using AutoML tools in real-world projects at a large scale. Therefore, we conducted an empirical study to understand how ML practitioners use AutoML tools in their projects. To this end, we examined the top 10 most used AutoML tools and their respective usages in a large number of open-source project repositories hosted on GitHub. 
The results of our study show 1) which AutoML tools are mostly used by ML practitioners and 2) the characteristics of the repositories that use these AutoML tools. Also, we identified the purpose of using AutoML tools (e.g. model parameter sampling, search space management, model evaluation/error-analysis, Data/ feature transformation, and data labeling) and the stages of the ML pipeline (e.g. feature engineering) where AutoML tools are used. Finally, we report how often AutoML tools are used together in the same source code files. We hope our results can help ML practitioners learn about different AutoML tools and their usages, so that they can pick the right tool for their purposes. Besides, AutoML tool developers can benefit from our findings to gain insight into the usages of their tools and improve their tools to better fit the users' usages and needs.
\end{abstract}


\input{textfiles/Introduction}

\input{textfiles/Background}
\input{textfiles/related_works}

\input{textfiles/experimental_setup}

\input{textfiles/results}
\input{textfiles/discussion}
\input{textfiles/threads_to_validity}

\input{textfiles/conclusion}
\input{textfiles/ack}
\balance
\bibliographystyle{IEEEtran}
\bibliography{reference.bib}

\end{document}

%% file: textfiles/Introduction.tex
\section{Introduction}
Due to the growth of available data, the usage of machine learning (ML) has increased dramatically over the past few years and a lot of business communities  and researchers are using ML to analyze their data and reach their goals~\cite{elshawi2019:automated}. ML practitioners usually have to follow a set of repetitive tasks including 
data collection and data preprocessing, feature engineering, hyperparameter optimization, model training, evaluating the performance of the resulting models, and deploying and monitoring the models in the production environment. As a result, a new field of automated machine learning (AutoML) has emerged over the past few years to automate and reduce the effort and time consumed by 
these repetitive tasks~\cite{truong2019:towards}, such as model training/ retraining and hyperparameter optimization in the ML pipeline.
For example, Tpot is an AutoML tool that uses genetic algorithm to optimize the ML pipeline. In addition, a lot of AutoML tools such as  Autosklearn~\cite{feurer-neurips15a,feurer-arxiv20a}, Autokeras \cite{jin2019auto}, Snorkel~\cite{snorkel:2021}, and Optuna~\cite{akiba2019:optuna} have been introduced for automating model management, data labeling, and optimizing hyperparameters. 

The popularity of AutoML tools has attracted the attention of researchers~\cite{xin2021:whither,wang2021:much,drozdal2020:trust}. For instance, Xin et al~\cite{xin2021:whither} conducted a qualitative study to understand the use of AutoML tools in practice by examining users' experience when using AutoML tools, the integration of AutoML features in 
ML workflows and the challenges faced 
when using the features of the AutoML tools in ML workflow. They suggested that AutoML tools should support the interactive user interface design to explore the intermediate execution (i.e., ``human-in-loop''). 
Also, through a qualitative study, Wang et al.~\cite{wang2021:much} studied how much automation is required by data scientists and reported that designing a ``human-in-the-loop'' AutoML tool is better than a tool that Completely automates the ML works~\cite{wang2021:much}. Drozdal et al.~\cite{drozdal2020:trust} studied what impacts the trust in the ML models built from using AutoML tools, through qualitative studies. They found that the non-functional requirements such as model performance metrics, visualization, and transparency 
increase the confidence of data scientists in AutoML tools. 

Although prior works performed qualitative studies to understand ML practitioners' experiences of using AutoML tools, it remains unclear how real-world projects use AutoML tools in their ML pipelines. Our study attempts to complement the existing studies through large-scale analysis of real-world open-source projects that use AutoML tools, aiming to get new empirical evidence and insights on the use of AutoML tools. Specifically, we extracted the information of the projects that use AutoML tools from GitHub and followed a mixture of both qualitative and quantitative analyses. We organize our study through the following research questions (RQs): 

    \nd $\bullet$ \textbf{RQ1: What are the most used AutoML tools?}
    We examined the usage of AutoML tools in GitHub projects and we identified the top 10 most used AutoML tools: Optuna, HyperOpt, Skopt, Featuretools, Tpot, Bayes\_opt, Autokeras, Auto-sklearn, AX, and Snorkel. Then, we analyzed the characteristics of the projects that use these tools, including their development history and popularity. ML practitioners can learn from our findings about which AutoML tools are mostly used and their characteristics when searching for 
    the right tools for their ML pipeline. 
    Besides, our findings can help AutoML tool developers gain insights into the projects that use their tools, which could allow them to make better decisions to further improve their tools. 
    
    \nd $\bullet$ \textbf{RQ2: How do ML practitioners use AutoML tools?}
    Through manual analysis, we examined 
    the purposes of using AutoML tools and the stages of the ML pipeline where they are used. We observed that AutoML tools are used mainly during the Hyperparameter optimization, Model training, and Model evaluation stages of the ML pipeline. 
    We also observed 10 main purposes of using AutoML tools, such as Hyperparameter optimization, Model management, Data management, and Visualization. ML practitioners can save time and effort by using AutoML tools to automate their ML pipeline tasks~\cite{truong2019:towards} for similar purposes in similar stages of ML pipeline.
    
    \nd $\bullet$ \textbf{RQ3: Are different AutoML tools used together?}
    We analyzed the source code of the projects using AutoML tools to examine whether AutoML tools are used together in the same source code files. 
    We found that, among the top 10 AutoML tools, only a few of them were used together, including Hyperopt being used together with other tools, such as Tpot, Skopt, Optuna. 
    Our findings can help ML practitioners with the needs for heterogeneous AutoML features choose the combination of tools that are mostly used together by other ML practitioners. In addition, developers of AutoML tools can leverage our findings to 
    provide APIs to make the collaboration between their tool and other co-used tools easier.

\nd \textbf{Organization:} The rest of the paper is organized as follows. In Section \rom{2}, the background about the AutoML tools and ML pipeline is provided. In Section \rom{3}, we provided a summary of the important related works. In Section \rom{4}, 
the experimental setup of the study is explained. The results of the research questions are presented in Section \rom{5}. A discussion of these results is presented in Section \rom{6}. 
Section \rom{7} discusses threats to the validity of our study. 
In Section \rom{8}, we conclude the paper and outline some avenues for 
future works.

%% file: textfiles/Background.tex
\section{Background}
AutoML tools are used to automate one or more stages (e.g., hyperparameter optimization) in the ML pipeline. This section provides background information on a typical ML pipeline and AutoML tools. In this article, the word ``ML practitioner'' refers to anyone who uses AutoML or ML in their project or work. An ML practitioner can be a developer, a researcher, a data scientist, or an ML engineer.

\subsection{Machine learning pipeline}
\begin{figure}[ht!]
\center
\includegraphics[width=\linewidth]{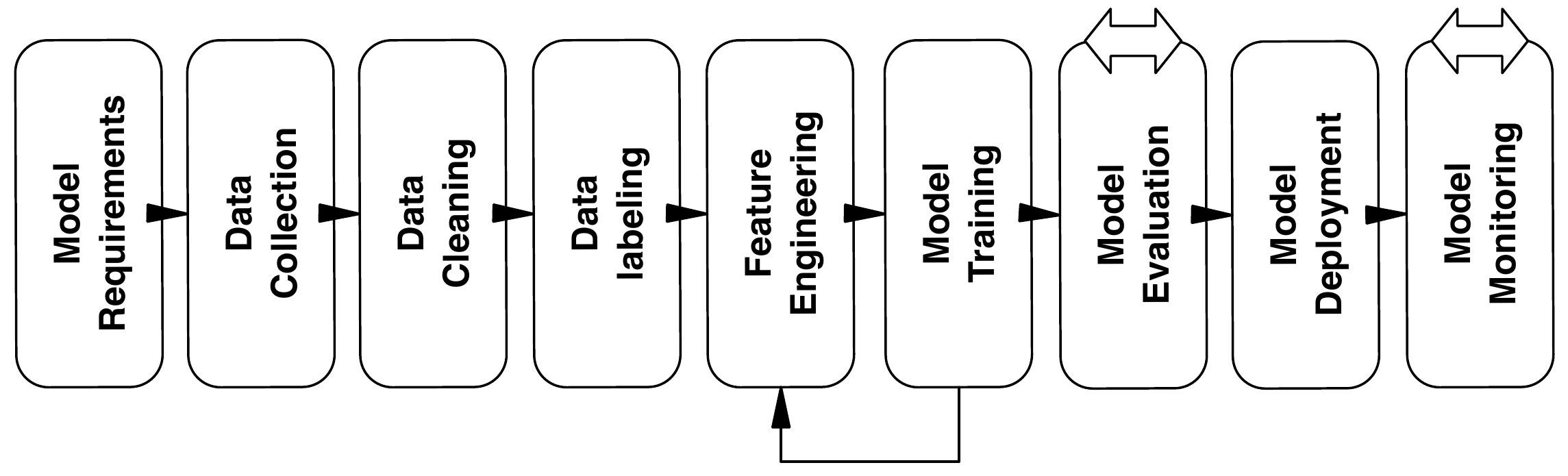}
\caption{The stages of the ML pipeline based on ~\cite{amershi2019:software}.}
\label{fig:The_Stages_of_ML_pipeline}
\vspace{-2mm}
\end{figure}

    Figure~\ref{fig:The_Stages_of_ML_pipeline} shows the nine stages of a typical ML pipeline for building and deploying ML software systems used at Microsoft, described by Amershi et al.~\cite{amershi2019:software}. We consider it as a reference ML pipeline in our study. Similar ML pipelines are also adopted by other commercial companies, such as Google~\cite{MLOps:Google:2021}, or IBM~\cite{IBM:MLOps:20210}. In the following, we describe the different stages of the ML pipeline mentioned in~\cite{amershi2019:software}. 
    
    The \emph {model requirements} stage is where ML practitioners select the type of ML model that fits their problem and products. 
    In \emph {data collection} stage, ML practitioners identify the data from different sources to collect, such as database systems or file storage. 
    The stage of \emph {data cleaning} includes cleaning the data to remove any anomalies that would likely hinder the training of the ML model. 
    The stage of \emph {data labeling} refers to assigning the proper labels to the data, such as assigning labels to a set of images reflecting the name of the objects present in the image. 
    
    In the stage of \emph {feature engineering}, ML practitioners prepare (e.g., transforming or normalizing the features/data) and validate the features that should be used in the model training phase. The \emph {model training} phase refers to training the ML models using the prepared features and the different implemented ML algorithms. 
    The result from this stage is a trained ML model used to predict the label of new incoming data. 
    In the \emph {model evaluation} stage, the predictive performance of the trained ML models is evaluated on a validation dataset. 
    
    The \emph {model deployment} stage is where the evaluated ML model and the entire workflow are exported for deployment. 
    Finally, The \emph {model monitoring} stage is where the model behavior is observed in the production environment to ensure that the model is doing what is expected to do.
    
\subsection{Automated machine learning (AutoML)}
The primary goal of automation in ML started from reducing 
human effort in the loop, specifically during the hyperparameter tuning and model selection stage~\cite{yao2018:taking}. However, the success of hyperparameter tuning and model selection also depends on other steps, such as data cleaning and preparation or feature engineering~\cite{wang2019:human, yao2018:taking,zoller2021:benchmark,crisan2021:fits}. More recently, AutoML has been broadly used in automating multiple steps of the machine learning workflow, from data collection and preparation to the deployment and monitoring of the ML model in production.

A variety of AutoML tools have been proposed to automate the computational work of building an ML pipeline. Some of these tools are built upon widely used ML frameworks. For example, Auto-Sklearn~\cite{feurer2020:auto, feurer2015:efficient}, and TPOT~\cite{olson2016:tpot, olson2016:evaluation} are built on  scikit-learn~\cite{pedregosa2011:scikit}.
These AutoML tools focus largely on supporting data preparation~\cite{lee2020:human,yao2018:taking,zoller2021:benchmark}, feature engineering \cite{kanter2015deep}, hyperparameter tuning \cite{akiba2019optuna}, model selection ~\cite{yao2018:taking,zoller2021:benchmark,fernandez2014:we}, and end-to-end development support for machine learning software systems~\cite{li2021:volcanoml}.

%% file: textfiles/related_works.tex
\section{Related Works}
The related research papers are categorized into two groups and discussed in this section.
\subsection{Qualitative studies on the usage of AutoML tools}
Prior work performs qualitative studies (e.g., through interviews) to understand practitioners' experiences of using AutoML tools~\cite{xin2021:whither,wang2019:human,crisan2021fits,wang2021:much,drozdal2020:trust,wang2019atmseer,xanthopoulos2020putting}.
For example, Xin et al.~\cite{xin2021:whither} investigated the use of AutoML tools through semi-structured interviews with AutoML users of different categories and expertise. 
They argued that although AutoML tools increase productivity for experts and make ML accessible to beginners, 
designing completely automated ML tools is impractical, and designers of AutoML tools should focus on improving the collaboration between ML practitioners and AutoML tools. 
Similarly, Wang et al. \cite{wang2021:much} 
argued that there is no need for tools that are fully automated, instead, developing AutoML tools that are explainable and that integrate people in the ML workflow loop is more important.
Crisan et al.~\cite{crisan2021fits} 
observed that using AutoML tools in real practices is not easy and suggested that one way to facilitate the interaction between humans and AutoML tools is data visualization. 
Similarly, Drozdal et al.~\cite{drozdal2020:trust} 
found that visualization and inclusion of transparency features build trust in users. Also, they recommended that AutoML tools should be easily customized for enabling users to apply their preferences.
Xanthopoulos et al.~\cite{xanthopoulos2020putting} defined some criteria to evaluate the features of AutoML tools.  
They concluded that the performance does not guarantee the success of an AutoML tool because there are other important factors that should be taken into consideration, such as the ease of use and the interpretability of the results. 

\subsection{Comparison of AutoML tools}
Ferreira et al.\cite{ferreira2021:comparison} studied eight AutoML tools and compared them by conducting computational experiments based on General Machine Learning (GML), deep learning, and XGboost scenarios. The results show that current GML AutoML tools can produce close or even better predictive results than human-created ML models.
Truong et al.~\cite{truong2019:towards} reported the different stages of the ML pipeline covered by some specific AutoML tools. Then, they examined the performance of the AutoML tools on different datasets. They found that none of the studied AutoML tools perform better than others. They suggest that AutoML tool developers should consider leveraging ideas from other AutoML tools to improve their tools. 

The previous works studied 1) the usage of AutoML tools by conducting qualitative studies such as interviews, 
2) the comparison between AutoML tools. However, no work studied how ML practitioners use AutoML tools in real-world projects at a large scale. In this work, we attempt to fill this gap by investigating the usage of AutoML tools in real-world GitHub projects. We aim to provide insights for ML practitioners and AutoML tool providers to improve the usage and design of AutoML tools. 

%% file: textfiles/experimental_setup.tex
\section{Experimental Setup} 

\begin{figure}[ht!]
\center
\includegraphics[width=\linewidth]{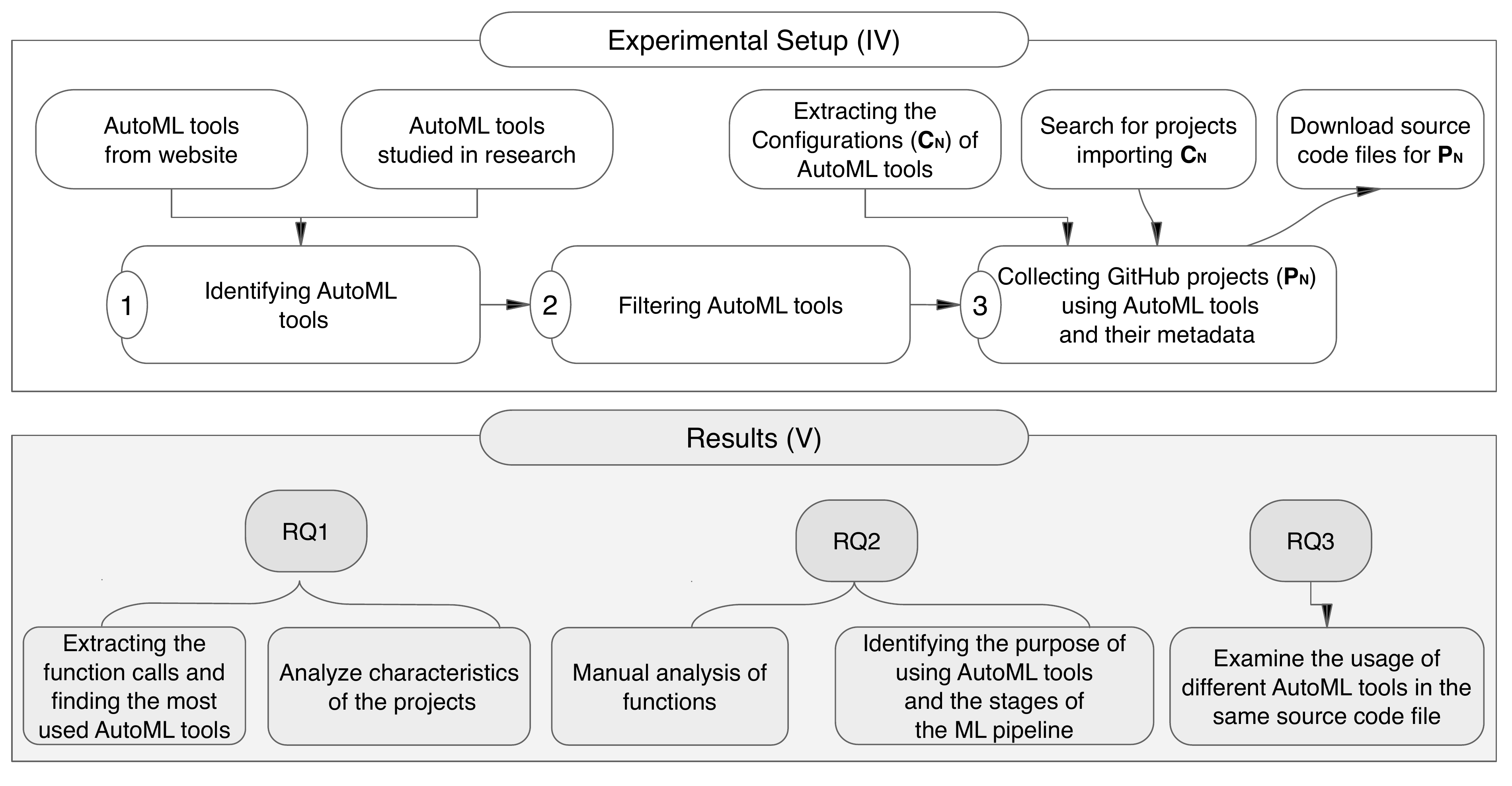}
\vspace{-4mm}
\caption{An overview of our study design. \textbf{\emph{Cn:}} configurations of a given AutoML tool (e.g., "import+tpot"), \textbf{\emph{Pn:}} GitHub project that are using AutoML (i.e., containing the source code file that imports atleast one of the AutoML configuration), \emph{IV} and \emph{V} correspond to the Section indices.} 

\label{fig:methodology}
\vspace{-5mm}
\end{figure}

This section describes the methodology followed to answer the proposed research questions. Our study design consists of a mixture of qualitative and quantitative (i.e., a sequential mixed-methods~\cite{ivankova:2006:using}) approaches. Figure~\ref{fig:methodology} show an overview of our study design explained in the following paragraphs.

\input{textfiles/data_collection}

%% file: textfiles/data_collection.tex
\textbf{\circled{1} Identifying AutoML tools:}

In this initial step of our methodology, the AutoML tools were identified from two sources: research papers and Google search. The two sources allow us to cover a broader range of AutoML tools proposed in the research publications and industry. In the following, we describe the identification steps from each data source.

$\bullet$ \textit{AutoML tools from the research papers:} 

To extract the list of AutoML tools mentioned in research papers. We used the search keywords ``automated'' AND ``machine learning'' to extract the scholarly literature related to AutoML on the Google scholar~\cite{Googlescholar:2022}
and Engineering Village (using Inspect and Compendex databases)~\cite{Engineeringvillage:2022}
platforms. We limited our search to the literature published between 2010 to 2021. Then, we examine the first 10 research papers from each of the two sources to collect the list of AutoML tools, as the search results of both sources are sorted based on the papers' relevance with the search keywords. Specifically, we were interested in the open-source AutoML tools that host their source codes on GitHub to allow us to study in detail their usage. In total, we extracted \textbf{31} open-source AutoML tools from the research papers.

$\bullet$ \textit{AutoML tools from Google search:} 

We used Google search to identify the names of open-source AutoML tools mentioned across different websites. We tried different keywords and finally settled for the keywords: ``(automl OR ``automated machine learning'') AND tools AND (GitHub OR list OR curated)'' and extracted 
all the possible AutoML tools from the first 100 resulting websites. We created a list of open-source AutoML tools that have a GitHub repository in the next step. One of the results was the awesomeopensource~\cite{awesomeopensourceo} website that contains 426 AutoML open-source projects. However, extracting all the possible AutoML tools listed on the website requires significant manual effort. Therefore, to get the most out of these tools listed, we limited our criteria to the most popular AutoML tools listed on the awesomeopensource~\cite{awesomeopensourceo} website by considering the AutoML tools with more than 1,000 stars. In total, we collected the names of \textbf{95} AutoML tools from Google search. 

\textbf{\circled{2} Filtering AutoML tools:}

After extracting the lists of names of AutoML tools from research papers and Google search results, we merged the two lists and removed the duplication. Then, using GitHub API~\cite{github:api} we extracted the repository details of each of these AutoML tools on November 24, 2021. For each AutoML tool's repository, we collected the following metadata: the number of contributors, the number of releases, description, number of forks, size of the repositories, creation date, last update, fork status, programming languages, number of stars, number of commits, and last commit date. We removed the tools with only one contributor (8 tools removed) or zero stars (1 tool removed). Also, using the last update date of the tools, we removed the tools that have not been active  
since 2019 or earlier (12 tools removed) because we are interested in the tools that are still under development due to the growth of the technology. Also, we manually checked all the remaining tools to verify that they are real AutoML tools. At the end of this step, we remained with \textbf{57} AutoML tools.

\textbf{\circled{3} Collecting GitHub projects using AutoML tools:}
\label{section:experimental-setup-1}

To find the projects that use open-source AutoML tools, we collected the configurations of each AutoML tool identified in the previous step by examining their documentation, GitHub repository, and sample source code. For example, \texttt{`import+optuna'} and \texttt{`from+optuna'} are the configurations that we found for Optuna. We only focused on the tools for which the configurations are in the Python programming language because Python is known as one of the main programming languages for building ML models and ML software projects~\cite{Python_language1:2017,Python_language2:2021}. Therefore, we removed the tools for which the configurations are not written in Python and the tools for which we could not find the configurations in their documentations (3 tools removed).

Next, using the AutoML configurations as keywords, we searched GitHub API for the source codes and the respective GitHub projects that import the configuration. For example, we searched for the keywords \texttt{`from+tpot'} and \texttt{`import+tpot'} in GitHub API to find repositories and the corresponding source codes that use the Tpot AutoML tool.
Then, we collected the paths to 97,815 source code files that contain at least one configuration of an AutoML tool and the name of the corresponding repositories (33,568 repository names collected) on December 4th and 5th, 2021. In the third step, we removed duplication from the resulting repositories (10,985 duplicates removed) and the source code files (39,194 duplicates removed).

Some repositories that use AutoML tools are not real projects, such as the course assignment repositories. We are not interested in these projects because they may not reflect the actual use of AutoML tools in real projects. Therefore, we identified the keywords that may reflect non-real projects. To do so, we randomly sampled 378 repositories from our dataset with a 95\% confidence level and 5\% confidence interval. We manually analyzed these repositorires and identified keywords includeing “assignment”, “book”, “chapter”, “tutorial”, and “course” (case insensitive). We used these keywords and removed the repositories with at least one of these keywords appearing in their name (602 repositories removed). Also, we removed the corresponding source code files path of the not-real projects (2,568 source code file paths removed). Furthermore, following the best practices established by previous studies~\cite{munaiah:2017:curating,Businge:2018:ICSME,Businge:2019:SANER,businge2022:reuse,openja2022:docker}, we removed the repositories and the corresponding source code file paths of the forked projects. This step ensures that the studied projects are not copies of another project but the mainline projects (190 source code file paths removed).
Finally, \textbf{21,981} projects and \textbf{55,863} source code file paths are retained in our dataset. 
Then, using GitHub API, we extracted each repository's metadata, such as the number of contributors, number of commits, description, forks, repository size, creation date, update date, last commit date, fork status, and number of stars. We collected the repositories' details on December 5th and 6th. Finally, we downloaded the source code files using the collected paths on December 8, 2021, for further analysis.

The rest of the data analysis is described in the results Section~\ref{sec:results}. We share the replication package of this study in~\cite{replication:AutoML-empirical-study:2022}.

%% file: textfiles/results.tex
\section{Results}\label{sec:results}

\input{textfiles/RQ1}

\input{textfiles/RQ2}

\input{textfiles/RQ3}

%% file: textfiles/RQ1.tex
\subsection{\textbf{RQ1: What are the most used AutoML tools?}}
\subsubsection{\textbf{Motivation}}

Although AutoML tools help ML experts and non-expert practitioners save effort and time by automating the repetitive tasks \cite{truong2019:towards}, choosing the right AutoML tool is still a challenging task~\cite{xin2021:whither,passi2018trust,wang2019:human,drozdal2020:trust}. The goal of RQ1 is to highlight the most used AutoML tools and provide detailed information about the projects using these tools. This information can help ML practitioners select appropriate AutoML tools for their projects. 
Besides, the insights gained from the projects using the tools can help AutoML tool's developers improve their tools in future releases. 

\subsubsection{\textbf{Approach}}
To find the most used AutoML tools, we collected the AutoML function calls in the GitHub projects. Also, we analyzed the characteristics of the GitHub projects that are using AutoML tools.

\nd \textbf{Collecting function calls of AutoML tools in GitHub projects:}
To understand the popularity of the AutoML tools, we analyzed the source code files of the GitHub projects that use AutoML tools and collected the calls that invoke the functions of each AutoML tool. 
Usually, a function is designed to perform a specific kind of task, whereby the name of that function often reflects that task. Therefore, calling such a function within a code invokes the tasks specified in the function. Invoking more functions of an AutoML tool indicates the tool is more engaged in a project. Thus we use the number of function calls to represent the popularity of an AutoML tool.
The following steps are taken in order to extract the function calls from the source code files:
Step 1) All Python notebooks files are converted to python file using the \emph{nbconvert} \cite{nbconvert:2022} and \emph{nbformat} \cite{nbformat:2022} Python libraries.
Step 2) The Python2 source code files are converted to Python3. Step 3) The syntax errors are resolved by removing the lines of source code files that start with \emph{!}, \emph{@}, \emph{\%}, \emph{\$}, \emph{install}, \emph{pip}, \emph{conda}. Step 4) The imported AutoML tools' libraries and their corresponding imported functions are extracted from each source code file using the abstract syntax tree (AST) Python library \cite{ast:2022}. Step 5) The AST Python library is used to extract the function calls related to the imported AutoML tools' libraries and the corresponding imported functions from the source code files. Step 6) The functions that are:a) directly mentioned in the code, b) called in the right side of an assignment, c) called in  the if, while, for loop, try, with, list, assert, return, tuple, unary operation statements, d) super functions of a class in which the parent is an AutoML tool are extracted. After collecting the function calls from all source code files, we counted the total number of calls of the functions of each AutoML tool. Then, we sorted the AutoML tools based on the number of total function calls among all projects in our dataset.

\textbf{Analyzing the characteristics of GitHub projects using AutoML tools:}
To understand the characteristics of the projects that use AutoML tools, we analyzed the details of the projects that use AutoML tools, such as the number of contributors, number of forks, size, age, star, and commits.

\subsubsection{\textbf{Results}}\label{result:RQ1}
\begin{figure}[ht!]
\vspace{-2mm}
\center
\includegraphics[width=9cm, height=4.5cm]{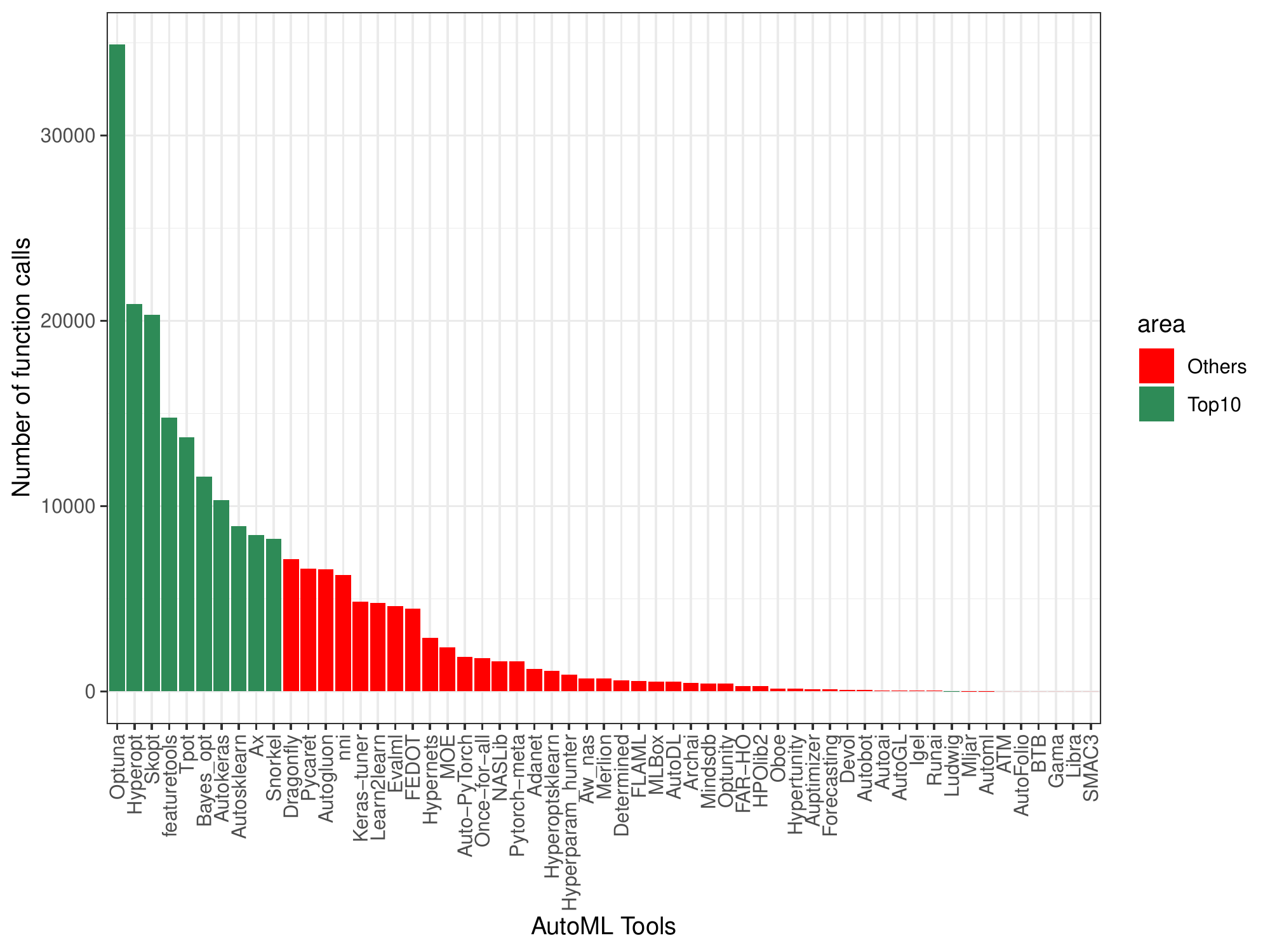}
\vspace{-4mm}
\caption{AutoML tools and the number of function calls.}
\label{fig:RQ1-Scope1}
\vspace{-12pt}
\end{figure}

Figure~\ref{fig:RQ1-Scope1} shows the total number of function calls for each AutoML tool. As seen in Figure~\ref{fig:RQ1-Scope1}, the top 10 AutoML tools with the most number of function calls are specified in green color. The percentage of function calls that are associated with the top 10 most used tools (i.e., 18\% of the 57 selected AutoML tools) is 68\%; these top tools include Optuna, HyperOpt, Skopt, Featuretools, Tpot, Bayes\_opt, Autokeras, Auto-sklearn, AX, and Snorkel. The details of each of these top 10 most used AutoML tools are provided in Table~\ref{tab:image-summary}. As seen in Table ~\ref{tab:image-summary}, the most used AutoML tools are popular based on the number of forks and number of stars. Also, they are 67.1 months old in average, which means that they are not young AutoML tools and it took them years to become popular.\par

\begin{tcolorbox}
The 18\% (10 out of 57) most used AutoML tools (e.g., Optuna) account for 68\% of all function calls. 

\end{tcolorbox}

\begin{figure}[ht!]
\center
\includegraphics[width=0.5\textwidth]{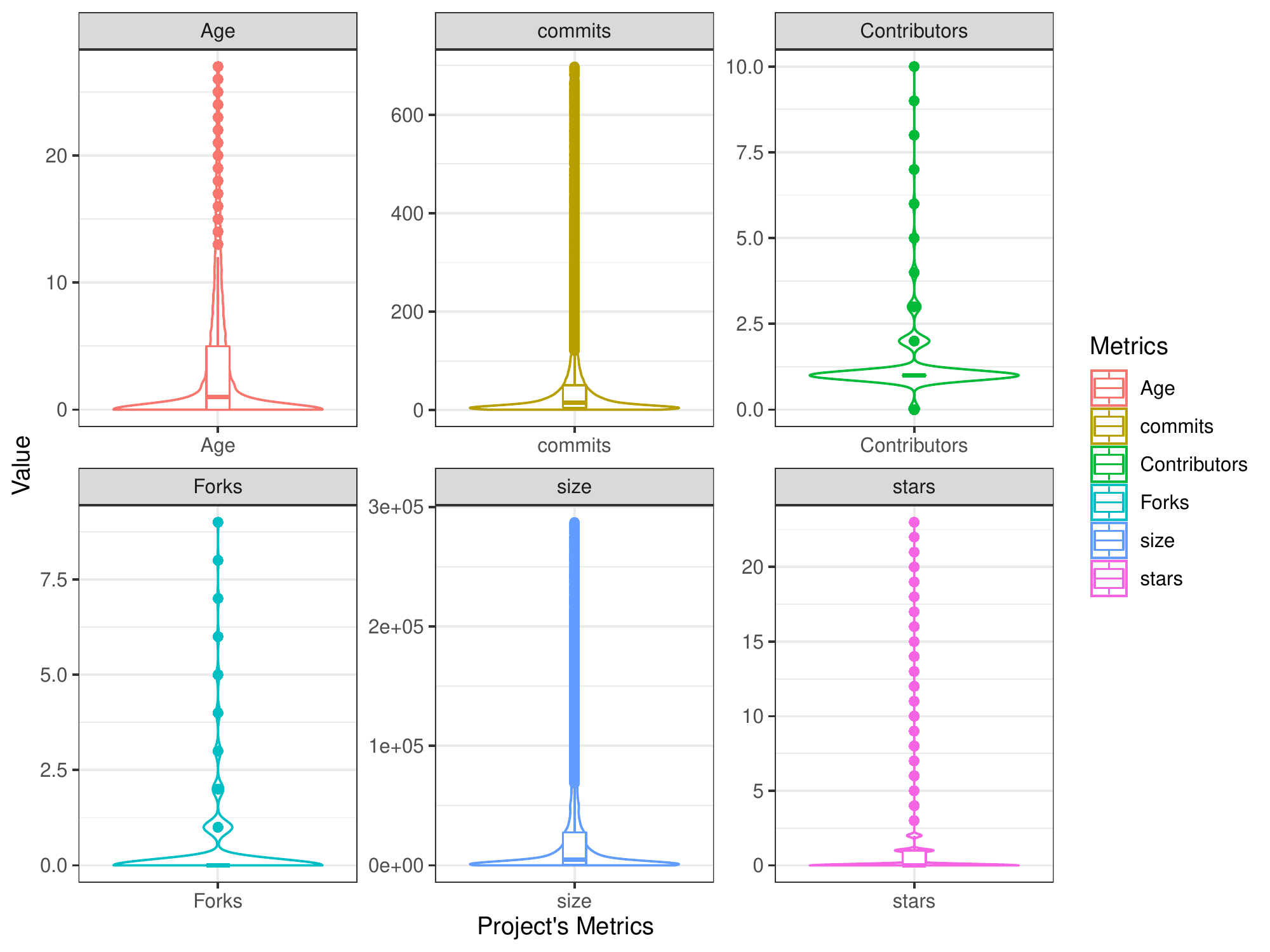}
\vspace{-4mm}
\caption{The summary characteristics of the projects that use AutoML tools in terms of their total Age in months (i.e., the difference between the creation date and the latest commit date), number of commits, number of contributors, number of forks, project's size based on lines of code, and the number of stars in each project.}
\label{fig:the summary of characteristics}
\vspace{-4mm}
\end{figure}

Furthermore, the profile of the projects that are using AutoML tools is shown in Figure~\ref{fig:the summary of characteristics}. For clear visualization, we removed the outliers within the last 5 percent of data in each metric and showed the distribution of 95\% percent of the data in Figure \ref{fig:the summary of characteristics}. As seen in Figure ~\ref{fig:the summary of characteristics}, 75\% of the repositories that are using AutoML tools are less than five months old and have fewer than 65 commits, two contributors, two forks, 35,753 lines of code, two stars. On the other hand, the projects that use AutoML tools also include some projects that are very mature (up to 10 
years in age), popular (with up to 102,983 
stars), with many contributors (with up to 433 
contributors), and large in size (with up to 17,453,577 
lines of code). 

\begin{tcolorbox}
The most used AutoML tools are popular tools according to their number of stars and forks. Furthermore, most of the projects that use AutoML tools are young and not popular (based on the number of stars).
\end{tcolorbox}

\input{tables/tbl_top_automl}

%% file: tables/tbl_top_automl.tex
\begin{table*}[t]
    \small
    \caption{\textbf{Summary statistics of the top 10 AutoML tools.}\\ 
    \emph{Function:} \emph{Number of function calls}, 
    \emph{S-code:} \emph{Number of unique source files importing the tool.} 
    \emph{Project:} \emph{Number of unique projects importing the tool}, 
    \emph{Size:} \emph{Size of the tool.} 
    \emph{Stars:} \emph{Total number of stars.}
    \emph{Age:} \emph{Age of the tool (i.e., the difference between latest commit date and the creation date).}
    \emph{Forks:} \emph{Number of forks associated to the tool.} 
    \emph{Releases:} \emph{The total number of releases associated to the tool.}
    \emph{Descriptions:} \emph{The description of the tool.}}
    \label{tab:image-summary}
   
   \begin{adjustbox}{width=1\linewidth,center}
    \begin{tabular}{r|l | l| c c c c c  c c  l}\toprule

     \rowcolor{gray!15}
    &\textbf{AutoML repository (short name)}&	\textbf{Function}&	\textbf{S-code}&	\textbf{Project}&	\textbf{Size}&	\textbf{Stars}&	\textbf{Age}&\textbf{Forks}&\textbf{Release}&\textbf{Descriptions}\\\midrule
1&optuna/optuna (Optuna)&34,916&5,482&3,140&13,577&5,553&45&602&41&A software framework for automatic hyperarameter optimization\cite{Optuna:2022}\\ 
\rowcolor{gray!15}
2&hyperopt/hyperopt (Hyperopt)&20,903&5,271&4,432&6,059&5,955&122&925&0&A library to optimize hyperparameter\cite{Hyperopt:2021}\\ 

3&scikit-optimize/scikit-optimize (Skopt) &20,313&2,807&1,830&9,423&2,238&68&424&23&"Sequential model-based optimization" \cite{Scikit-optimize:2021}\\ 
\rowcolor{gray!15}
4&Featuretools/featuretools (Featuretools)&14,786&1,053&548&5,888&5,864&50&769&94&A library to automate feature engineering\cite{Featuretools:2022}\\ 
5&EpistasisLab/tpot (Tpot)&13,725&2,893&1,796&7,9030&8,349&72&1,437&27&A library to optimize machine learning pipeline\cite{tpot:2021}\\ 
\rowcolor{gray!15}
6&fmfn/BayesianOptimization (Bayes\_opt)&11,582&3,128&2,605&23,290&5,543&89&1,222&7&"Global optimization with gaussian processes" \cite{BayesianOptimization:2020}\\ 
7&keras-team/autokeras (Autokeras)&10,331&1,259&550&44,487&8,241&48&1,334&53&"An AutoML system based on Keras"\cite{autokeras:2022}\\ 
\rowcolor{gray!15}
8&automl/auto-sklearn (Auto-sklearn)&8,909&974&543&7,4993&5,870&76&1,092&29&An AutoML toolkit based on sklearn \cite{auto-sklearn:2022}\\ 
9&facebook/Ax (AX)&8,442&722&213&283,337&1,649&33&178&23&A adaptive platform for adaptive experiments \cite{ax:2022}\\ \rowcolor{gray!15}
10&snorkel-team/snorkel (Snorkel)&8,237&853&264&292,468&4,922&68&797&14&A tool for generating training data \cite{snorkel:2021}\\  

\bottomrule
    
    \end{tabular}
    \end{adjustbox}
\vspace{-6mm}
\end{table*}

%% file: textfiles/RQ2.tex
\subsection{\textbf{RQ2 : How do ML practitioners use AutoML tools?}}
\subsubsection{\textbf{Motivation}}
ML practitioners may use AutoML tools for different purposes across their ML pipeline. We analyzed the stages of the ML pipeline in which AutoML tools are used and the purposes of using AutoML tools to help AutoML tools' developers gain more insights on the usage of their tool and enable them to improve their tools in a more efficient way. 

Understanding the stages of the ML pipeline in which the AutoML tools are most popular can help developers of the tools find the shortcomings of their tools and help improve their tools efficiently.
Besides, our results can provide insights for ML practitioners to better leverage AutoML tools in different stages of their ML pipeline and for different purposes. 
\vspace{2mm}

\subsubsection{\textbf{Approach}}

To understand the purposes of using AutoML tools and the stages of the ML pipeline where they are being used,
we used the extracted function calls in Section ~\ref{result:RQ1} and labeled them. To reduce the required time and effort for manual labeling, we only focused on the function calls of the top 10 most used AutoML tools (identified in RQ1).
Besides, we focused on the function calls that are most commonly used in different projects. 
Thus, from the set of all unique function calls of the selected AutoML tools, we compute their total frequency and Shannon entropy value, then choose the functions that have more than 80\% of 1) the Shannon entropy value and 2) the total frequency for manual labeling. In the following, we explain 1) the process of selecting functions based on Shannon entropy and usage frequency and 2) the process of manual labeling.

\nd \textbf{Selecting functions based on Shannon entropy and usage frequency}:
We used the normalized Shannon's entropy~\cite{shannon:2001mathematical,khomh:2011entropy} to compute the distribution of the function calls across the projects: 
   
   \begin{equation}\label{Shannon entropy equation}
    H_{n}(F)=-\sum_{i=1}^{i=n} p_{i}*\log_{n}(p_{i})
    \end{equation}
    In Equation (\ref{Shannon entropy equation}), \emph{F} is a function that we want to measure its Shannon's entropy value, \emph{n} is the total number of unique projects, \emph{i} is the unique number of projects,
     \emph{$p_{i}$} is the probability of the use of function  $F(p_{i}\geq 0$, and $\sum_{i=1}^{i=n} p_{i} = 1)$ in a given project $i$.    
     $p_{i}$ is calculated by dividing the total number of occurrence of function \emph{F} in the project \emph{i} by the total number of occurrence of function \emph{F} in all the projects.
     If all the unique projects have the same probability of using function \emph{F} (i.e., $p_i = \frac{1}{n}, \forall i \in 1,2,..,n$), the Shannon entropy of \emph{F} is maximal  (i.e., $H_{n}(F)=1$).
    Thus, the function \emph{F} with higher Shannon entropy is more likely to be used more by ML practitioners compared to the functions with less Shannon entropy.
    
    The functions with $H_{n}(F) > 0.8$ and with their respective counts of function calls above 80\% percentage (sorted in descending order) were chosen for manual labeling. A total of 619 functions from the top 10 AutoML tools were selected in this step.
    
\nd \textbf{Manual labeling}: Here we try to understand why AutoML tools are used by ML practitioners and what they are used for. To this end, we manually labeled the selected functions in the previous step and formed categories and sub-categories of purposes of using AutoML tools. We also assigned a separate label representing the stages of the ML pipeline for which the respective functions are called, following the stages presented in Figure~\ref{fig:The_Stages_of_ML_pipeline}. The first two authors were involved in the initial labeling and construction of the taxonomy. Both individuals are graduate students with strong knowledge in machine learning and software engineering. The primary reference during the labeling is the official documentation of the functions from the tool websites and the publications associated with the AutoML tools. For example, the functions \texttt{`autokeras.TextClassifier.evaluate'}~\cite{autokeras3:2022}
and  \texttt{`autokeras.ImageRegressor.evaluate'}~\cite{autokeras2:2022}
of AutoKeras are used to evaluate the best model and their purposes were labeled as `Model evaluation'. 
The functions related to data formatting, generating statistics of the input data, or cleaning is assigned the stage of `Data cleaning'.

During the first iteration of labeling, the first author labeled the functions of Optuna, Skopt, Featuretools, and Ax, and the second author labeled the functions of the remaining six tools. In the Second iteration, all the initial labels were combined, and the first two authors underwent through each label together and subsequently further discussed the labels that were not agreed upon until a consensus was achieved. 
A random sample of 238 functions was selected with a confidence level of 95\% and a confidence interval of 5\% 
for further validation by the other two authors (not involved in the initial labeling). They agreed on 98\% of the stages of the functions and 96\% of the purposes of the functions.
\subsubsection{\textbf{Results}}
Table \ref{label of stage of ML-pipeline}
details the stage of the ML pipeline (i.e., the table columns) in which ML practitioners use AutoML tools. Each cell value in Table~\ref{label of stage of ML-pipeline} represents the percentage of the total number of calls of the functions from the respective AutoML tool corresponding to the stages of the ML pipeline. As shown in Table~\ref{label of stage of ML-pipeline}, the AutoML tools are mostly used for hyperparameter optimization (accounting for 40.9\% of the function calls). One of the possible reasons is that hyperparameter optimization involves multiple repetitive tasks and requires a lot of effort and time~\cite{truong2019:towards}; similarly, there might be numerous different sub-tasks involved compared to other stages of the ML pipeline.

\begin{tcolorbox}
AutoML tools are mainly used ($40.9\%$) during the Hyperparameter Optimization stage of the ML pipeline, followed by Model training ($11.5\%$), Model evaluation ($11.2\%$), and Feature Engineering ($10.1\%$). AutoML tools mostly focus on providing functions for training ML models effectively, selecting a model, and evaluating the performance of the resulting model.
\end{tcolorbox}


\input{tables/tbl_stage_composition_rq2}

\begin{figure*}[ht!]
\center
\includegraphics[width=0.80\linewidth]{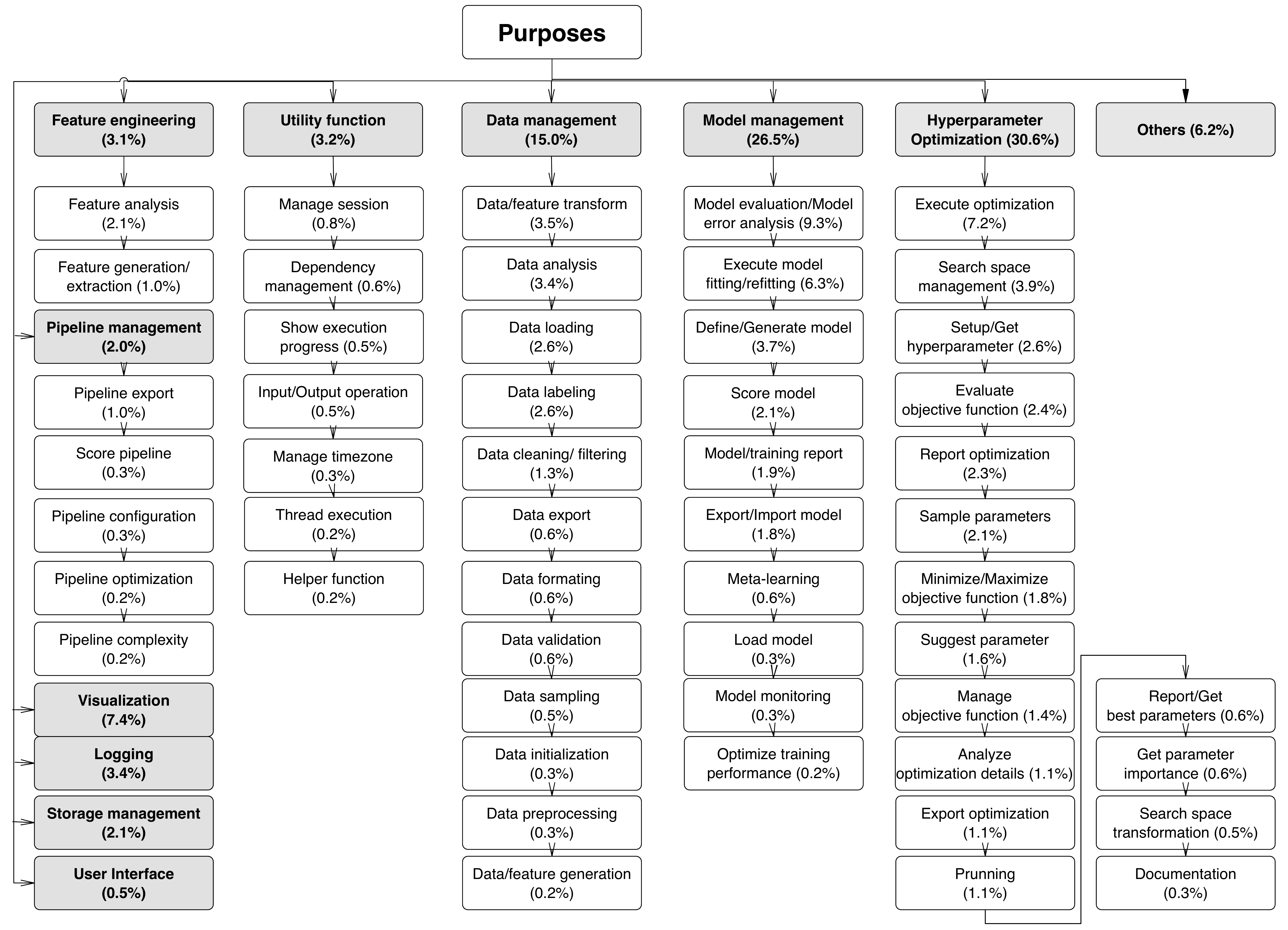}
\caption{The taxonomy of the purposes of using AutoML tools. 
The high level categories are highlighted in grey, while the sub-categories are in white boxes.}
\label{fig:automl-functions-taxonomy}
\vspace{-5mm}
\end{figure*}

Figure \ref{fig:automl-functions-taxonomy} presents high-level (in the grey background) and low-level (in white background) categories of the purposes for which  ML practitioners use AutoML tools. As shown in Figure \ref{fig:automl-functions-taxonomy}, the high-level categories of purposes are in the order: Hyperparameter Optimization (30.6\%), Model management (26.5\%), Data management (15\%), Visualization (7.4\%), Logging (3.4\%), Utility functions (3.2\%), Feature Engineering (3.1\%), Storage management (2.1\%), Pipeline management (2\%) and User Interface (0.5\%). The following section describes some of the categories and sub-categories of purposes.

\nd $\bullet$ \textbf{Hyperparameter optimization (HPO):} This category is about using AutoML to determine the right combination of hyperparameters that maximizes the model performance, usually through running multiple experiments in a single training process. Automating HPO is indeed crucial, particularly for Deep Neural Networks (DNN)~\cite{feurer2019:hyperparameter, akiba2019:optuna}, which depend on a wide range of hyperparameter choices about the function tuning or regularization, parameter sampling, the network’s architecture, and optimization. In Figure~\ref{fig:automl-functions-taxonomy}, we reported $16$ different activities involved during the HPO process, from executing hyperparameter optimization and parameter sampling to HPO documentation. 

\nd $\bullet$ \textbf{Model management:} This category is about using AutoML to handle the different tasks related to training ML algorithms with some input data and the resulting model that can generate predictions using patterns in the input data.  In Figure~\ref{fig:automl-functions-taxonomy} we highlighted the different sub-tasks where ML practitioners use AutoML in the ML models management, using AutoML for model evaluation and error-analysis, model training/testing, model definition, model monitoring, and optimizing model training performance.

\nd $\bullet$ \textbf{Data management:} In this category, AutoML is used to handle the tasks related to the data used for building ML models. This process includes a set of steps involving data, including data generation and processing steps, managing data flow to ML models, working with multiple datasets, and other tasks throughout the data pipeline. In particular, we summarized $12$ different sub-tasks related to data management in Figure~\ref{fig:automl-functions-taxonomy} including data transformation, data analysis, data loading, data labeling, and data export.

\nd $\bullet$ \textbf{Utility functions:} These are functions provided by specific AutoML tools to perform some common functionally and are often reusable in the ML pipeline. The Utility functions are particularly important to reduce the pipeline complexity and increase the readability of the source code. Examples of the tasks in this category include session management (e.g., during model training), handling of input/output operations of ML models, showing training or optimization progress, and handling of thread execution.

\nd $\bullet$ \textbf{Feature engineering:} This category is about using AutoML tools to automate the extraction of features from the given data, usually using domain knowledge of the data and the model being built. The resulting features are used as input to some algorithms to train the ML models. We reported in Figure~\ref{fig:automl-functions-taxonomy} two main sub-categories of Feature engineering, including feature analysis and feature generation/extraction.

\nd $\bullet$ \textbf{Pipeline management:} This category is about using AutoML to automate the end-to-end management of an ML model or a set of multiple models, including the orchestration of data that flows into or out of the model.

\nd $\bullet$ \textbf{Visualization:} This category is about the graphical representation of data and information, such as visualizing the feature importance using a chart for interpretability or visual representation of data distribution during data analysis.
    
\nd $\bullet$ \textbf{Logging:} This category involves using AutoML tools to track and store records related to the execution events, data inputs, processes, data outputs, resulting in model performance and other related tasks when developing and deploying ML models. ML practitioners use the resulting information to identify any suspicious activities in ML software projects' privileged operations and assess the impact of state transformations on performance.

\nd $\bullet$ \textbf{Storage management:} In this category, AutoML tools are used for managing the data storage resources aiming at improving and maximizing the efficiency of data storage resources, often in terms of performance, availability, recoverability, and capacity.

\nd $\bullet$ \textbf{User interface:} This category is about the element of AutoML tools that allows the ML practitioners to easily interact with the AutoML features, using the concepts such as visual and interactive design and information architecture (e.g., providing web-based interactive interface for displaying live code).

We used the label \textbf{Others} for the function calls which do not fall into any of the categories.


\begin{tcolorbox}
We derived 10 high-level categories of the purposes of using AutoML tools, including Hyperparameter optimization (30.6\%), Model management (26.5\%), Data management (15\%), Visualization (7.4\%), and Logging (3.4\%). ML practitioners can save time and effort by using AutoML tools to automate their ML pipeline tasks~\cite{truong2019:towards} for similar purposes.
\end{tcolorbox}

\begin{figure}[ht!]
\center
\includegraphics[width=\linewidth]{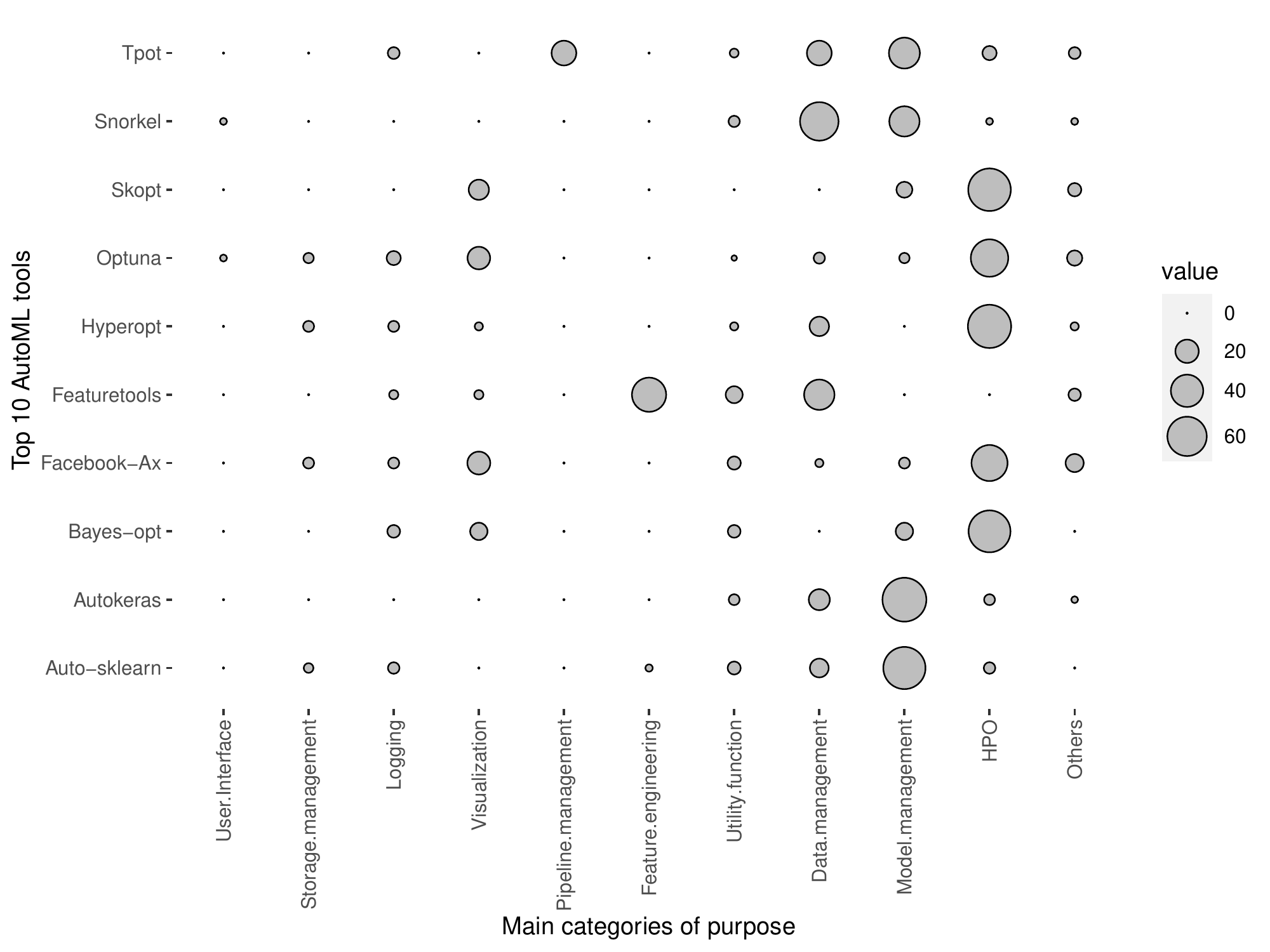}
\vspace{-2mm}
\caption{The percentage composition of the high-level categories for purposes of using AutoML across the top 10 most used tools.}
\label{fig:purpose-tools-correlation}
\vspace{-5mm}
\end{figure}

In Figure~\ref{fig:purpose-tools-correlation} we provide the breakdown of the major categories of purposes (x-axis) of using AutoML provided in the top 10 AutoML tools (y-axis). The circle size represents the composition of using the respective AutoML tools for that purpose, computed as the percentage of total function calls defining that purpose to the total number of function calls from the individual AutoML tool.  

We can clearly see in Figure~\ref{fig:purpose-tools-correlation} that the functions for Hyperparameter optimization, Model management, and Data management are used the most for most of the top 10 AutoML tools. Specifically, Hyperparameter optimization and Model management dominate at least 40\% of the AutoML tools. In contrast, some of the purposes of using AutoML tools are specific to certain tools. For example, we can observe that the function calls for Feature engineering ($45\%$) and Pipeline management ($22.7\%$) tasks are dominantly used from Featuretools and Tpot, respectively, while a minimal percentage ($1.3\%$) of function calls related to User Interface comes from Snorkel and Optuna.

%% file: tables/tbl_stage_composition_rq2.tex
\begin{table}[!t]
    \centering
    \small
    \caption{The percentage usage of AutoML functions across the ML workflow. \\ 
    }
     \label{tab:automl-tools-stages}
   
   \begin{adjustbox}{width=1\linewidth,center}
    \begin{tabular}{r|m{1.5cm} m{1.5cm} m{1.5cm} m{1.5cm} m{1.5cm} m{1.5cm}  m{1.5cm} m{1.5cm} m{1.5cm} m{1.5cm} }\toprule

     \rowcolor{gray!15}
    \textbf{AutoML}&	\textbf{Data collection}&	\textbf{Data cleaning}&	\textbf{Dala labeling}&	\textbf{Feature engineering}&	\textbf{Hyperparameter optimization}&	\textbf{Model training}&\textbf{Model evaluation}&\textbf{Model deployment}&\textbf{Model monitor}&\textbf{Others}\\\midrule
    
    Optuna&1.4&2.9&0&0&72.9&2.1&0.7&0.7&12.1&6.4\\
    \rowcolor{gray!10}
    hyperopt&11.5&1.9&0&5.8&75&0&0&0&3.8&1.9\\
    
Skopt&0&0&0&0&90.6&3.1&3.1&0&0&3.1\\
\rowcolor{gray!10}
Featuretools&12.8&7.7&0&69.2&0&0&0&0&2.6&7.7\\

Tpot&0&11.6&0&9.3&9.3&23.3&18.6&18.6&4.7&4.7\\
\rowcolor{gray!10}
Bayes-opt&0&0&0&0&78.9&0&0&0&5.3&15.8\\

AutoKeras&3.6&3.6&0&8.3&8.3&40.5&25&7.1&0&3.6\\
\rowcolor{gray!10}
Auto-sklearn&2.7&5.4&1.4&5.4&4.1&25.7&44.6&1.4&4.1&5.4\\

Facebook-Ax&0&4.2&0&0&68.8&4.2&4.2&2.1&4.2&12.5\\
\rowcolor{gray!10}
Snorkel&9.5&21.6&27&2.7&1.4&16.2&16.2&1.4&0&4.1\\

\bottomrule
{AVERAGE}&{4.2}&{5.9}&{2.8}&{10.1}&\textbf{40.9}&\textbf{11.5}&\textbf{11.2}&{3.1}&{3.7}&{6.5}\\

\bottomrule
    
    \end{tabular}
    \end{adjustbox}
\label{label of stage of ML-pipeline}
\vspace{-5mm}
\end{table}

%% file: textfiles/RQ3.tex
\subsection{\textbf{RQ3: Are different AutoML tools used together?}}

\subsubsection{\textbf{Motivation}}
From Table~\ref{tab:automl-tools-stages} we observe that at least two of the AutoML tools can be used to automate the functionality in each stage of the ML pipeline. This question aims to understand whether the different AutoML tools are used together in the same source code file. For instance, ML practitioners may use AutoSklearn for model training and tune the hyperparameters using the functionality provided by Optuna within the same code. Answering this question could help ML practitioners choose the AutoML tools used together for different purposes. Similarly, the information can help the developer of the AutoML tools improve the integration procedure of the AutoML tools used together, such as providing an API to facilitate the usage of the other AutoML tools while getting input or sending output.

\subsubsection{\textbf{Approach}}


We compared the composition of the source code files (of the projects using AutoML tool) where different AutoML tools are used together as follows. For every source code files $\{Ts_{1}, Ts_{2}, Ts_{3}, ... Ts_{n}\}$ importing and using the functions from a given AutoML tool, we computed the percentage composition of the shared source files as:

\begin{equation}\label{equa-shared-files}
\frac{Ts_{i\cdot j}*100}{Ts_{iN}} 
\end{equation}
where $Ts_{i\cdot j}$ is the total number of source files using the functions of two AutoML tools $T_i$ and $T_j$ in the same file, and $Ts_{iN}$ is the total number of files using at least one function from the AutoML tool $T_i$. We define the results of the equation as the correlation between two AutoML tools.

In addition, we also calculated the distribution of the projects that use the different number of AutoML tools to understand how these AutoML tools are used together in the same projects.



\subsubsection{\textbf{Results}}
Table~\ref{tab:project-automl-automl} shows the distribution of the projects and source codes that use a different number of AutoML tools. We focus on the top 10 AutoML tools and the projects that use these tools. We observe that the vast majority of the projects use only one AutoML tool. There are only fewer than 8\% of the projects that use two or more different AutoML tools in the same project.

\begin{figure}[ht!]
\center
\includegraphics[width=\linewidth]{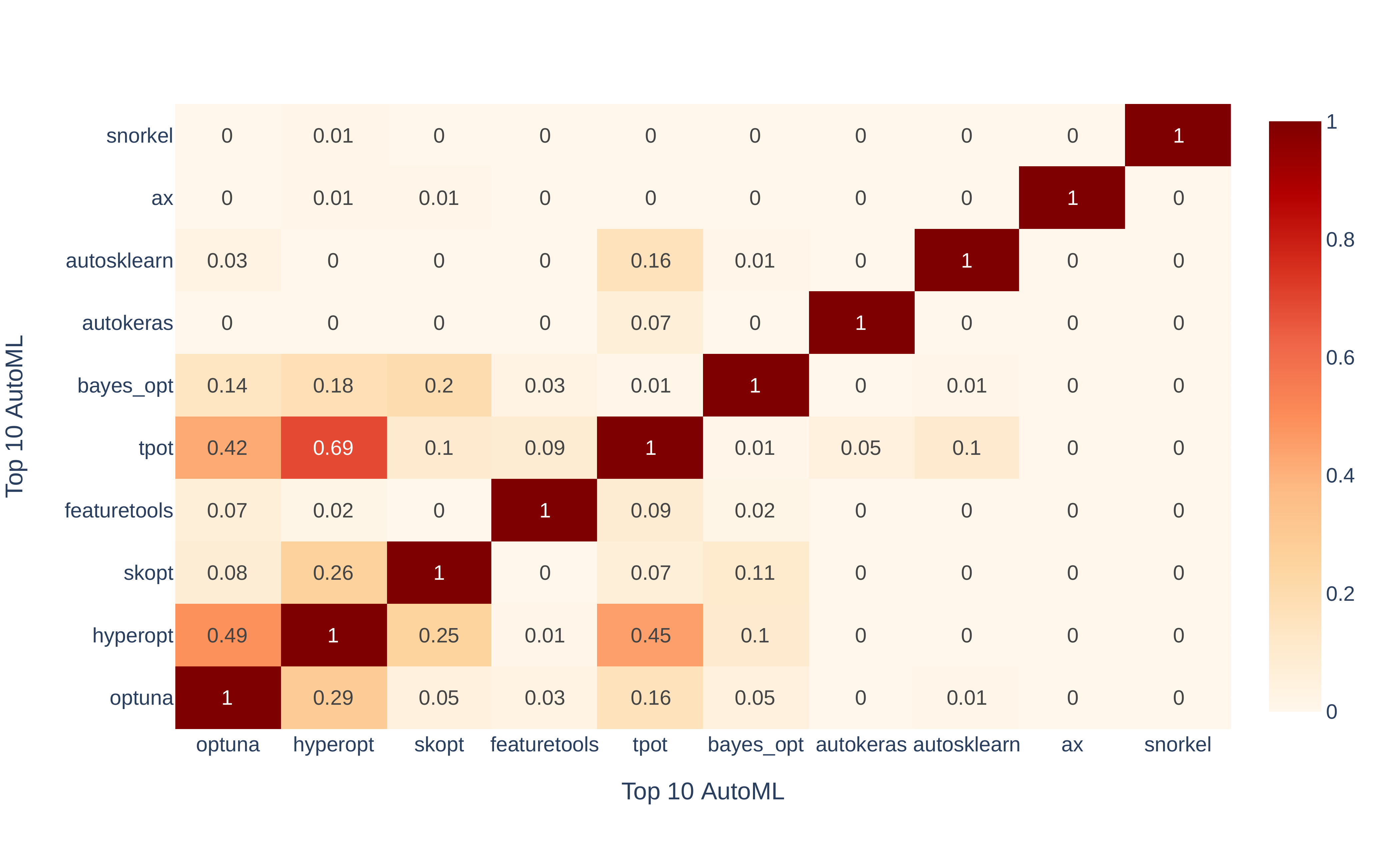}
\vspace{-4mm}
\caption{The composition of project's source code files using different AutoML tools.}
\label{fig:automl-correlation-matrix}
\vspace{-5mm}
\end{figure}

Figure~\ref{fig:automl-correlation-matrix} shows the correlation matrix 
of the percentage composition of shared source files using different top AutoML tools together, computed using Equation~(\ref{equa-shared-files}). We only reported the correlation of the top 10 most used AutoML tools for clear visualization.

\input{tables/tbl_projects_automl_counts}

From Figure~\ref{fig:automl-correlation-matrix} we observe that, across the top 10 AutoML tools, only a few of the source files ($<2\%$) 
uses the functions from more than one AutoML tool. This result implies that most of these tools are used independently. Among the most used AutoML tools, the few AutoML tools used together are Tpot with Hyperopt ($0.69\%$), 
Optuna with Hyperopt  ($0.29\%$), Skopt with Hyperopt ($0.25\%$) and Bayes-Opt with Skopt ($0.20\%$). 
In Table~\ref{tab:project-automl-function-example}, we provide examples of the functions used together in the same source code file. For instance, both Hyperopt and Optuna provide rich functions for optimizing hyperparameters according to Figure~\ref{fig:purpose-tools-correlation}; however, optuna also offers other features that are useful during hyperparameter optimization, such as Visualization and Logging. This implies that ML practitioners using HyperOpt with Optuna together have more options for HPO while also utilizing the visualization and logging functions from Optuna. 
Similarly, Tpot provides few functions for optimizing hyperparameters but offers more functions such as pipeline management and model management; the ML practitioners can simultaneously automate these tasks when using Tpot with Hyperopt. 




\begin{tcolorbox}
Less than 8\% of projects use two or more different AutoML tools.
Different AutoML tools are rarely used in the same source code files. Among the few AutoML tools used together, Hyperopt is used with other tools, such as Tpot, Skopt, and Optuna. ML practitioners can consider using the combination of functions from different AutoML tools together to help automate multiple tasks across the ML pipeline. 
Similarly, the AutoML developers can propose an efficient API to integrate the AutoML tools that are used together.
\end{tcolorbox}

%% file: tables/tbl_projects_automl_counts.tex
\begin{table}[t]
    \centering
    \small
    \caption{The percentage number of GitHub projects and source code files using different AutoML tools together.}
    \label{tab:project-automl-automl}
     \begin{adjustbox}{width=0.95\linewidth,center}
    \begin{tabular}{ l | m {2cm} m {2cm} m {2cm} m {2cm} m {2cm} m {2cm}}\toprule
    
    \textbf{No. AutoML}&\textbf{1}&\textbf{2}&\textbf{3}&\textbf{4}&\textbf{5}&\textbf{$\geq 6$}\\\hline
    
    \textbf{\% No.of Projects}&92.52&5.85&1.28&0.26&0.053&0.045\\
    
     \textbf{\% No.of code files}&98.77&0.90&0.32&0.004&0&0\\

    
    \end{tabular}
     \end{adjustbox}
\vspace{-6mm}
\end{table}

\begin{table}[t]
    \centering
    \small
    \caption{Examples of the AutoML functions used together .}
    \label{tab:project-automl-function-example}
     \begin{adjustbox}{width=0.95\linewidth,center}
    \begin{tabular}{ m {3cm} | m {10cm} }\toprule
    \textbf{Tools combination}&\textbf{Example function}\\\hline
     \rowcolor{gray!10}
     hyperopt, tpot&\texttt{hyperopt.fmin',
   'tpot.TPOTClassifier.fit', 'tpot.TPOTClassifier.score'}\\
   
   hyperopt, optuna& \texttt{'hyperopt.fmin',
   'hyperopt.hp.choice',
   'hyperopt.hp.uniform',
   'optuna.create\_study',
   'optuna.create\_study.optimize'}\\
    \rowcolor{gray!10}
   bayes\_opt, skopt&\texttt{'bayes\_opt.BayesianOptimization', 'bayes\_opt.BayesianOptimization. maximize',
   'skopt.BayesSearchCV',
   'skopt.BayesSearchCV.fit'}\\
   tpot, hyperopt, optuna&\texttt{'hyperopt.Trials',
   'hyperopt.fmin',
   'optuna.create\_study',
   'optuna.create\_study.optimize',
   'tpot.TPOTClassifier.fit', 'tpot.TPOTClassifier.score'}

    
    \end{tabular}
     \end{adjustbox}
     \vspace{-6mm}
\end{table}

%% file: textfiles/discussion.tex
\section{Discussion and Implication}

 
This section further discusses the implications of our findings for the research community, the ML practitioners, and tool developers.

\nd $\bullet$ \textbf{Researchers:} Our study provides a general view of the usage of AutoML tools by ML practitioners across the ML pipeline and highlights the most used AutoML tools. Our initial results indicate that most of the popular AutoML tools are used in relatively immature and less popular projects, with 75\% of the projects being developed by fewer than two contributors and being younger than five months old. Also, the result shows that the top 10 tools (18\% of the tools) account for 68\% of the usages. Future work can explore the factors that impact the adoption of the AutoML tools. 

Also, regarding the usage of the AutoML tools in the studied projects, our results indicate that most of the functions used by the ML practitioners are model-oriented, specifically in the Hyperparameter optimization (40.9\%) and model training (11.5\%) stages of the ML pipeline. These results indicate that the current AutoML tools focus more on providing effective model training and the resulting models and pay little attention to the data prepossessing process. Indeed data prepossessing has been the least focus of AutoML tools, in line with the previous works~\cite{truong2019:towards,Anaconda2022:State, yao2018:taking}, yet an ML model is as good as the quality of the data~\cite{breck2019:data}. Researchers can investigate the challenges involved in automating the data-oriented tasks of the ML pipeline for a better quality of the data. 

\nd $\bullet$ \textbf{AutoML Tool Developers:} The developers of the AutoML tools, on the one hand, can use our results to improve their tools to support the most used purposes and stages of the ML pipeline while keeping in mind the target users of their tools. They can also add more features to support the stages not currently supported such as data labeling. On the other hand, the developers of the least used AutoML tools can improve their tools by implementing or reusing the similar functionalities provided by the most popular AutoML tools.

We also observe that, although not frequent, some AutoML tools are used together in the same source files. For example, Hyperopt is one of the tools that is used with other tools, such as Tpot, Skopt, and Optuna. The developers of these tools can propose an efficient API for integrating these tools. Also, AutoML tools' developers can propose better tooling/frameworks integrated into their AutoML tools for improving the quality of the data. Moreover, we observed that the functionalities related to documentation and user interface (UI) are among the least used, indicating that the current AutoML tools may suffer from the level of adaption (and possible customization) by the ML practitioners of diverse skills and expertise. Xin et al.~\cite{xin2021:whither} in their study found that customization is one of the lowest-rated quality attributes of AutoML tools, as reported by practitioners. 
ML practitioners expressed the need for more customization. They claim that this would improve their trust and transparency in using the AutoML tools~\cite{xin2021:whither}. Therefore, AutoML tools' developers should consider providing more visualization functionalities and improving the UI, to increase the usability of their tool. 

\nd $\bullet$ \textbf{ML Practitioners:} We encourage the ML practitioners to use our findings to learn about the popular AutoML tools and the different purposes for which they can be used, in order to automate similar tasks in their ML pipeline; saving both time and effort~\cite{truong2019:towards}. Also, they can choose the combination of the tools that are mostly used together including Hyperopt with other tools, such as Tpot, Skopt, or Optuna. For instance, when the ML practitioner wants to optimize hyperparameters and manage the entire ML pipeline, Hyperopt (for hyperparameter optimization) with Tpot (for pipeline management) can be a good choice according to Figures~\ref{fig:purpose-tools-correlation} and \ref{fig:automl-correlation-matrix}.
 

%% file: textfiles/threads_to_validity.tex
\section{Threats to validity} 
We now discuss threats to the validity of our study.\\
\nd \textbf{Internal validity threats:}
We manually studied a sample of AutoML function calls to understand ML practitioners' purposes for using AutoML tools. However, our labeling results may not truly reflect the actual purposes of the developers. Future work can perform developer surveys to confirm our results. 
Besides, the process of manual labeling affects the results of this study. To reduce the impact of biased labels, two authors labeled the functions, and the other two authors reviewed the labels separately to validate them. Another possible threat is related to our use of function calls to determine the popularity of the AutoML tools. As different numbers of calls may combine to achieve the same goal across different AutoML tools, which may affect the results of this study. However, a function is usually designed to perform a specific task, as reflected by the name of that function. Therefore, we believe that this threat is minimal. In the future, we will analyze the relation cardinality between functions of different AutoML tools to understand how the studied tools handle the same situation with the same or different number of function calls. 

\nd \textbf{External validity threats:} 
The number of selected AutoML tools for manual labeling affects the generalizability of the results. To minimize this effect, we selected the 10 most used AutoML tools, which account for 68\% of the usage in the GitHub projects. We assumed that our findings derived from these 10 AutoML tools can represent how ML practitioners use AutoML tools and their main purposes for using them.

\nd \textbf{Construct validity threats:} 
To extract the function calls of AutoML tools in GitHub projects, we followed the official documentation of AST library.
However, it's still possible that we may have missed some function calls that are rarely used and hard to recognize.
We provide all our scripts and data in our replication package~\cite{replication:AutoML-empirical-study:2022}, to allow replicating our results.

%% file: textfiles/conclusion.tex
\section{Conclusion and future work}

This paper presents an empirical study on the usage of AutoML tools in open-source projects hosted on GitHub. As the initial step, we examined the most used AutoML tools and the details of the projects using the tools. We reported the top 10 most used AutoML tools and showed that most of the projects using the tools are young and less popular. 
Then, we manually investigated the function calls of ML tools to understand the purposes of using AutoML tools and their used stages in the ML pipeline. 
We showed that some of the purposes (such as User Interface, Logging, Feature Engineering, and Pipeline management) are only specific to
a few AutoML tools (e.g., Optuna, Tpot and Featuretool) compared to other purposes (e.g., Hyperparameter Optimization, Model management, Data management) which span across multiple AutoML tools. We also showed that most AutoML tools focused on model training and model selection but had few functions to automate the data-oriented stages of the ML pipeline. Moreover, we examined whether the different AutoML tools are used together in the same source code and observed that only a lower percentage of the files use different AutoML tools together in the same code. 
Our future work will investigate the features that are provided by the AutoML tools but not used by ML practitioners. 
In addition, we will quantitatively study the factors that impact the successful adoption of the AutoML tools in real-world ML projects.

%% file: textfiles/ack.tex
\section*{Acknowledgement}
This work is funded by the Fonds de Recherche du Québec (FRQ), Natural Sciences and Engineering Research Council of Canada (NSERC), and Canadian Institute for Advanced Research (CIFAR).